# RT-SRTS: Angle-Agnostic Real-Time Simultaneous 3D Reconstruction and Tumor Segmentation from Single X-Ray Projection

Miao Zhu, B.S.[1, #], Qiming Fu, M.S.[1, #], Bo Liu, Ph.D.[1,*], Mengxi Zhang[1], Bojian Li[1], Xiaoyan Luo, Ph.D.[1,*], Fugen Zhou, Ph.D.[1],

[1] *Image Processing Center, Beihang University, Beijing 100191, People's Republic of China*

[#] Miao Zhu and Qiming Fu contributed equally

*Corresponding Author:

    Bo Liu, Ph.D.

    Image Processing Center

    Beihang University, Beijing, 100191, P.R. China

    Email: bo.liu@buaa.edu.cn

    Xiaoyan Luo, Ph.D.

    Image Processing Center

    Beihang University, Beijing, 100191, P.R. China

    Email: luoxy@buaa.edu.cn

Last revised: March 28, 2024

**Abstract**

Radiotherapy is one of the primary treatment methods for tumors, but the organ movement caused by respiration limits its accuracy. Recently, 3D imaging from a single X-ray projection has received extensive attention as a promising approach to address this issue. However, current methods can only reconstruct 3D images without directly locating the tumor and are only validated for fixed-angle imaging, which fails to fully meet the requirements of motion control in radiotherapy. In this study, a novel imaging method RT-SRTS is proposed which integrates 3D imaging and tumor segmentation into one network based on multi-task learning (MTL) and achieves real-time simultaneous 3D reconstruction and tumor segmentation from a single X-ray projection at any angle. Furthermore, the attention enhanced calibrator (AEC) and uncertain-region elaboration (URE) modules have been proposed to aid feature extraction and improve segmentation accuracy. The proposed method was evaluated on fifteen patient cases and compared with three state-of-the-art methods. It not only delivers superior 3D reconstruction but also demonstrates commendable tumor segmentation results. Simultaneous reconstruction and segmentation can be completed in approximately 70 ms, significantly faster than the required time threshold for real-time tumor tracking. The efficacies of both AEC and URE have also been validated in ablation studies. The code of work is available at https://github.com/ZywooSimple/RT-SRTS.

## 1 Introduction

Lung cancer is currently the leading cause of cancer-related deaths worldwide. Radiotherapy is considered a primary treatment modality for cancer patients, with approximately 77% of lung cancer patients requiring it with evidence-based indications[1]. However, in the realm of lung cancer radiotherapy, the inherent motion of tumors caused by respiration can significantly affect the treatment accuracy[2]. Gated and tracking treatment techniques have been proposed to address this issue to improve treatment outcomes. Essential to these techniques is the real-time monitoring of the tumor during treatment[3], coupled with the desire to obtain time-resolved 3D images for precise assessment of the radiation dose received by the patient [4, 5].

Owing to its importance, motion management in radiotherapy has been studied for decades[6, 7]. The most commonly used and proven clinical techniques are marker and surrogate-based methods. However, they suffer from accuracy decline caused by marker migration or varying relationships between the surrogates and the target[5, 8]. Additionally, implanting markers complicate the radiotherapy protocol and may result in clinical complications such as pneumothorax[8]. As a result, extensive attention has been paid to the marker-less monitoring of tumors based on intraoperative images, especially kV X-ray projection, which is easily accessible via the integrated imaging systems of the treatment machine. However, due to its perspective characteristics, it is tough to locate the tumor position from X-ray projection directly, and it is considered "Holy Grail" in the field of image guidance for radiotherapy[9]. Moreover, the complexity of the task is further heightened by the demand for advanced volumetric-modulated arc therapy (VMAT), which is widely employed in lung cancer radiotherapy. Given that both the treatment head and imaging system rotate around the patient during treatment, it is possible to acquire only one X-ray projection at a certain angle. Consequently, in the context of VMAT motion management, the objective becomes highly intricate, necessitating the simultaneous real-time reconstruction of the 3D CT image and localization of the tumor position based on a single X-ray projection at an arbitrary angle.

Fulfilling this task was highly challenging before the rise of deep learning. Although registration-based methods can achieve simultaneous reconstruction and tumor location, they suffer from high computational complexity and low robustness. Recently, convolutional neural network (CNN) has been studied to address this issue, and promising results have been obtained owing to its powerful nonlinear modeling ability and high inference efficiency. Notably, studies have demonstrated that it is possible to reconstruct 3D CT images with adequate quality from 1~2 X-ray projections. However, current methods fail to fully meet the motion control requirements for VMAT radiotherapy of lung tumors. First, current methods can only reconstruct 3D images but cannot directly realize the real-time localization of tumors. Second, these methods have only been validated for fixed-angle imaging, i.e. training a CNN model for an angle, and thus, cannot be utilized in VMAT in which the treatment beam and X-ray imaging system rotate during beam delivery. Therefore, in this work, a novel patient-specific CNN imaging method is proposed for angle-agnostic Real-Time Simultaneous 3D Reconstruction and Tumor Segmentation (RT-SRTS). This method adopts the multi-task learning (MLT) strategy to integrate 3D imaging and tumor segmentation into one network and can achieve simultaneous 3D reconstruction and tumor segmentation in real-time. A new attention enhanced calibrator (AEC) module is proposed to fuse the extracted hierarchical features better with the imaging and segmentation branches. The segmentation accuracy is further improved using a new uncertain-region elaboration (URE) module. The efficacy of the proposed method in the angle-agnostic imaging mode, i.e. training a CNN model for all angles, is investigated and the results are compared with three state-of-the-art methods in both the fixed-angle and angle-agnostic imaging modes. Better results are obtained using the proposed method, demonstrating its advantage. In addition, the performance in locating tumors is noteworthy compared to other reported results.

In summary, the contributions of this paper are as follows:
1)  This pioneering effort explores simultaneous real-time 3D CT reconstruction and tumor segmentation from a single X-ray projection acquired at any given angle, an essential aspect of image-guided VMAT for lung tumors. Previous works have primarily focused on fixed-angle CT reconstruction and required training in a separate network for each angle, which cannot meet the requirements of image-guided VMAT.

2) By leveraging the principles of multi-task learning, this study introduces a CNN imaging model featuring dual branches and incorporates a multi-task learning loss with two components. The CNN imaging model effectively translates shared features derived from a 2D projection image of a perspective nature into two 3D images, a novel exploration in the field.

3) Attention-enhanced calibrator (AEC) module and uncertain-region elaboration (URE) module have also been proposed to enhance the accuracy of 3D reconstruction and tumor segmentation. While the AEC module serves as a novel feature and spatial dimension calibrators enhanced with self-attention mechanism, the URE module integrates the uncertainty fixation concept into simultaneous reconstruction and segmentation.

4) The method is thoroughly validated through a comparison with three state-of-the-art methods on fifteen cases. The results demonstrate improved reconstruction outcomes, and it is possible to simultaneously achieve accurate tumor localization in 70±3 ms.

**2 Related works**

This study focuses on CT reconstruction and tumor localization based on a limited number of X-ray projections. Historically, these two aspects have been addressed through distinct research avenues. Hence, this section examines the existing literature concerning both CT reconstruction and tumor localization, providing insights into the separate but interconnected developments in these domains.

**2.1   Online tumor location based on X-ray projection**

One method for locating tumors focuses on directly analyzing and processing X-ray projections[10-12], the accuracy of which depends on the visibility and clarity of the tumor in an image. When the tumor is relatively small or blocked by other high-density substances, localization accuracy is difficult to ensure[4].

Another method achieves online tumor location via matching or registering pretreatment 3D/4D CT with X-ray projection, which can theoretically obtain both the tumor position and time-resolved 3D CT images[9, 13-15]. However, the registration of 3D images with one X-ray projection is an ill-conditioned problem. It is necessary to simplify the motion model or use prior motion knowledge to ensure the robustness and feasibility of the derived motion. Principal component analysis-based motion model (PCA-MM) is one such method that has been successfully applied[16]. With PCA-MM, the registration problem is reduced to optimization of a few parameters, greatly reducing complexity and improving robustness[17]. However, iterative optimization is still required, which cannot meet the real-time requirements[18]. Additionally, these methods can only obtain a linear combination of a few principal motion components, making it challenging to handle variations in respiratory motion patterns or patient positions[19, 20].

Recently, convolutional neural network (CNN) has also been applied to data-driven registration between pre-treatment 3D/4D CT and X-ray projections. In one of the pioneering works, Wei et al. proposed to apply CNN to predict PCA-MM parameters from X-ray projections and achieved real-time 3D imaging and tumor location[5, 21-23]. In this method, a CNN regression model is trained to model the nonlinear mapping from the X-ray projection to the PCA-MM parameters during the pre-treatment training phase. Online X-ray images are input into the CNN model during treatment to obtain the PCA coefficients. The real-time CT image and tumor location are obtained by applying the resulting PCA motion model to the reference CT and tumor position. Under the similar framework, some recent works have proposed directly predicting the deformation field instead of using the PCA coefficient. For example, Dai et al. proposed the direct prediction of the deformation field of lung, which was used to deform the pre-treatment segmentation map[24]. This work was validated only for fixed-angle conditions and it was found that the performance at 0° (Left-right direction) was worse than that at 90° (Anterior-posterior direction).

However, its performance for angle-agnostic condition remains unknown. Shao et al. predicted the deformation field of a live surface and utilized biomechanical modeling to generate intra-liver deformation[25-27]. Biomechanical modeling is critical for accurate liver tumor location because of the low contrast between the tumor and liver tissue. However, solving for tumor motion using finite element analysis compromises the processing efficiency[24].

## 2.2  CT reconstruction from sparse measurement

The reconstruction of CT images from sparse measurements is a longstanding pursuit. On the one hand, it aims to reduce both the acquisition time and radiation dose to the patient. On the other hand, specific scenarios, such as those encountered in VMAT motion control, present challenges in obtaining sufficient measurements. Numerous endeavors have been dedicated to this end, and one particularly fruitful approach is the model-based iterative reconstruction (MBIR) approach[28]. Based on the principles of compressed sensing or maximum a posteriori[29-31], MBIR achieves high accuracy in CT reconstruction from sparse samples by solving constrained optimization models. MBIR offers notable advantages, including data consistency enforced by the underlying physical model. However, these approaches face parameter-tuning challenges and extended execution times.

In tandem with the flourishing development of deep learning across various medical imaging domains[32, 33], the field of CT reconstruction is also undergoing a revolution propelled by deep learning techniques. Current methods can be categorized into several groups based on the role of deep learning in the imaging processes[34]. An extensively studied category is the image domain learning approach, in which the CNN is employed as a post processing step to mitigate artefacts in a preliminary reconstruction. Another widely investigated category is the unrolling or unfolding approaches, which leverage the expressive power of neural networks to approximate the iterative reconstruction process in a more efficient and end-to-end trainable manner[28, 35, 36]. In addition, domain transform approaches have recently garnered significant attention. This approach directly learns 3D images from measurement data by harnessing the abundant medical big data and the powerful computing capacity. Domain transform approaches elevate sparsity to unprecedented levels, thereby enabling the reconstruction of CT images from a single X-ray projection. This breakthrough enabled the reconstruction of 3D CT images during VMAT for dose verification.

Based on the dataset involved, the current efforts to learn 3D CT images from X-ray projections using deep learning can be categorized into two types. The first type focuses on motion control for radiotherapy, aiming to learn a patient-specific imaging model from planning 4D CT and applying it to predict 3D images during radiotherapy. A pioneering study by Shen et al. demonstrated that reconstruction results based on a single X-ray projection were comparable to those obtained using 10 X-ray projections, affirming the feasibility of 3D imaging from a single X-ray projection[20]. Subsequently, Lei et al. introduced a generative adversarial network with perceptual supervision[4], while Lu et al. incorporated a super-resolution module to enhance the reconstruction resolution[37]. By contrast, the second type seeks to train a population-based CNN imaging model from the CT images of a group of patients and predict the CT images for any patient. In an early attempt, Henzler et al. attempted to reconstruct the 3D anatomy of the skull from a single X-ray image by using a simple encoding-decoding CNN structure[38]. Subsequently, many efforts have been made to improve imaging capacity[39-42]. Ying et al. formulated the problem in a generative adversarial network framework and proposed a novel skip connection module that naturally bridges 2D and 3D feature maps more naturally[39]. Tan et al. added an attention mechanism and a multi-scale feature fusion module to improve feature learning, and proposed a New Inception module[40, 41].

Despite the above progress in the field of CNN based imaging, it is noteworthy that existing methods predominantly focus on CT reconstruction without considering tumor segmentation, thus failing to meet all requirements for VMAT motion control. An exception is the work of X-CTRSNet [42], which considered both the reconstruction and segmentation of the cervical vertebra. However, in this approach, a separate segmentation network is concatenated with a reconstruction network to perform segmentation, which is straightforward but computationally extensive with two separate feature extraction modules. To address this gap, this study proposes a new patient-specific imaging method that couples reconstruction and segmentation within one network based on the MLT theory. It achieves simultaneous real-time 3D reconstruction and tumor segmentation from one X-ray projection and thus fully meets motion control requirements for VMAT radiotherapy.

## 3 Material and methods

### 3.1 Angle-agnostic real-time simultaneous reconstruction and tumor segmentation network

The proposed method aims to reconstruct a 3D CT image and segment the tumor simultaneously in real-time, based on an X-ray projection acquired at any angle. To achieve this, the task is formulated as a multi-task learning problem that has been used successfully across many machine learning applications. In addition to achieving multiple goals simultaneously, different tasks serve as regulations for each other, helping to learn more general representations and reducing the chance of overfitting. Following the common MTL strategy of sharing hidden representations, RT-SRTS is designed with a shared representation network and two decoding sub-networks. As shown in Figure 1, the shared representation network is utilized to extract hierarchical semantic features from the input 2D X-ray projection and transform the learned 2D features into a 3D format to achieve the transformation task from 2D to 3D. Based on this, reconstruction and segmentation sub-network were integrated to reconstruct the 3D CT image and generate the tumor mask in one network simultaneously.

A residual-learning scheme was applied in the representation network[43], in which a 2D residual block was used to facilitate the training process and avoid gradient vanishing during back-propagation. Five 2D convolution residual blocks with different number of convolutional filters were concatenated to learn hierarchical semantic representations from the 2D projections. Instance normalization was adopted as the training batch size is set to 1 because of memory limitations.

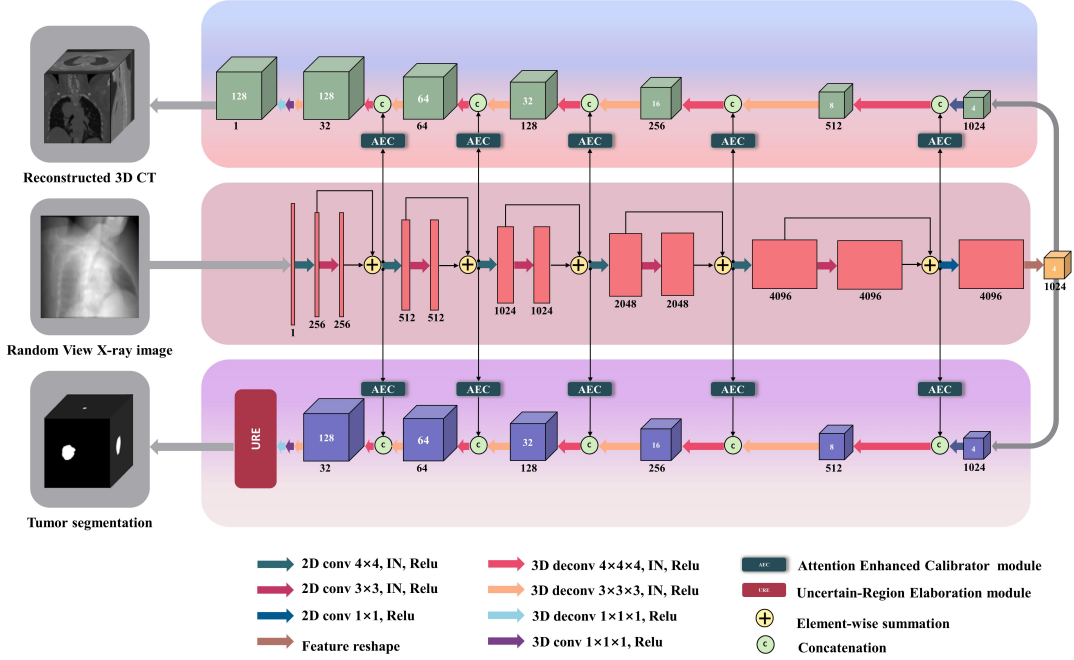

**Figure 1** The framework of the proposed method.

The reconstruction sub-network was constructed using a 3D deconvolution block. The first deconvolution layer up-samples the feature map by a factor of 2 and reduces the number of feature maps by decreasing the number of deconvolutional filters. The second deconvolution layer undergoes a transformation and maintains the spatial shape of the 3D features. In accordance with the representation network, there are five concatenated deconvolutional blocks, and a novel attention enhanced calibrator (AEC) module with attention strategy is proposed to utilize the hierarchical information from the representation network optimally.

The tumor segmentation sub-network shares a similar architecture to the CT reconstruction sub-network, except for the last few layers, where a SoftMax activation function is applied on a 2-channel feature to generate the initial segmentation results.

Subsequently, the URE module is applied at the end of the segmentation branch to obtain fine-grained segmentation.

### 3.1.1 The attention enhanced calibrator (AEC) module

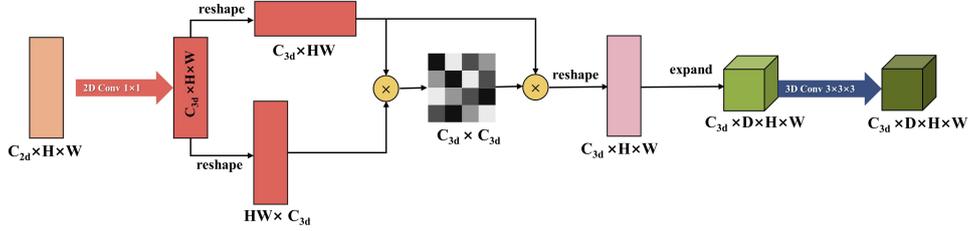

**Figure 2** The architecture of the AEC module.

The AEC module was designed as a special skip connection to transform hierarchical features from the representation sub-network to reconstruction and segmentation branches. In addition to allowing for an efficient flow of gradients for learning, the AEC performs two additional calibration functions. One is to calibrate the feature dimensions, and the other is to calibrate the spatial dimensions. After calibration, the output feature had consistent dimensions in both feature and spatial domains with the concatenated 3D feature in the two 3D branches.

As shown in Figure 2, for feature dimension calibration, a basic 2D convolution block was first applied to the input feature to compress the number of features from $C_{2d}$ to $C_{3d}$. Subsequently, a self-attention mechanism is applied to capture the relative relationships of various organs and tissues hidden within the features[44], aiming to obtain more representative features. Specifically, the correlation between features at different channels is calculated to obtain the inter-channel relationship of features, and the enhanced features with calibrated feature numbers ($C_{3d} \times H \times W$) are generated by taking the correlation weighted sum of all features. Finally, for spatial dimension calibration, to maintain computing efficiency, a simple but effective dimension expansion was applied to copy and stack 2D feature map ($C_{3d} \times H \times W$) to obtain 3D feature map ($C_{3d} \times D \times H \times W$).

### 3.1.2 The uncertain-region elaboration (URE) module

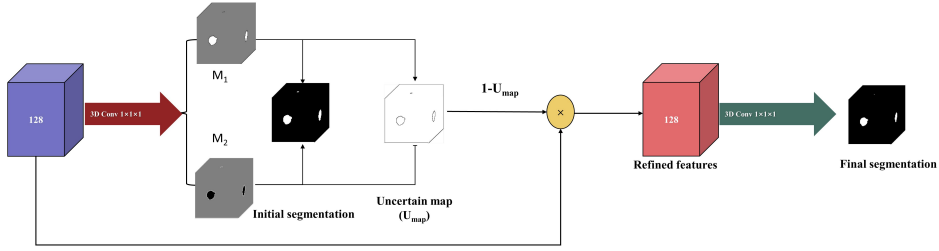

**Figure 3** The uncertain-region elaboration (URE) module.

The initial tumor segmentation results usually exist blurry and uncertain regions, particularly around the tumor boundaries. Inspired by recent works on uncertainty fixation[45], a novel module called, uncertain-region elaboration (URE), was proposed to reduce ambiguity and improve segmentation results. It includes a segmentation evaluation of such an uncertainty and a segmentation adjustment to obtain fine-grained results. The core idea of URE is to leverage pixels with high confidence to rectify pixels with low confidence. Figure 3 shows the detailed structure of the URE module. After decoding and SoftMax normalization, two probability maps, $M_1$ and $M_2$, were generated to indicate the likelihood of each pixel $p$ belonging to the tumor and the background, respectively. The value of $M_1$ and $M_2$ lies in [0,1], and $\sum_{i=1}^{2} M_i(p) = 1$. The label of $p$ is determined by comparing $M_1$ and $M_2$, and a pixel-wise uncertainty map, $U_{map}$, was generated to measure the classification ambiguity of each pixel:

$$U_{map} = 1 - \exp(1 - \frac{\max(M_1, M_2)}{1 - \max(M_1, M_2)}) \tag{1}$$

The $U_{map}$ has a value range of [0,1), with a larger $U_{map}$ value representing higher confidence. Once the uncertainty map was calculated, the initial features used for segmentation were multiplied by the $U_{map}$ and a local convolutional operator was added to generate refined features used to obtain the final segmentation.

3.1.3 Loss function and implementation details

According to the MLT, the loss function is designed in two parts: MSE loss between the generated and ground truth CTs and binary cross entropy (BCE) loss between the generated and ground truth tumor masks.

$$L_{loss} = \alpha_1 L_{MSE} + \alpha_2 L_{BCE} \quad (2)$$

$$L_{MSE} = \frac{1}{M \cdot N \cdot K} \sum_{i=0}^{M-1} \sum_{j=0}^{N-1} \sum_{m=0}^{K-1} (I(i,j,m) - T(i,j,m))^2 \quad (3)$$

$$L_{BCE} = \frac{1}{M \cdot N \cdot K} \sum_{i=0}^{M-1} \sum_{j=0}^{N-1} \sum_{m=0}^{K-1} -\left[ y \cdot \log p + (1-y) \cdot \log(1-p) \right] \quad (4)$$

where $M$, $N$, and $K$ represent the size of 3D CT image, $I$ and $T$ are the generated and ground truth CT images, $p$ represents the probability that a pixel is predicted as tumor, and $y$ represents the true label of the pixel with a value 0 stands for background. $\alpha_1$ and $\alpha_2$ are the weights for the reconstruction and segmentation losses. Choosing weights for different tasks in MLT is a crucial aspect that can significantly affect the performance of the overall model. The weights determined the importance of each task during the learning process. In RT-SRTS, because both tasks are of similar importance and complexity, a simple and straightforward approach was used to set the weights for both losses to 1.0, treating each task as equally important.

The RT-SRTS training steps are presented in Algorithm 1. It was implemented using PyTorch library and trained on a NVIDIA GeForce RTX 3090 GPU with 24 GB of memory. For comparison, the same training strategy and hyper-parameters were used for all experiments. The network was trained for 100 epochs in all patient cases, and the learning rate was initially set to $2e^{-3}$ and linearly decayed to zero from 50 to 100 epochs. Adam optimizer with $\beta_1 = 0.50$ and $\beta_2 = 0.99$ is used to minimize the total loss function and iteratively update network parameters through back propagation. At the end of each training epoch, the model was evaluated on the validation set and the best checkpoint model with the smallest validation loss was saved.

```
Algorithm 1: The training process of RT-SRTS
   Input:
 1 DRR images $X_{drr}$, corresponding 3D CT $Y_{ct}$ & tumor labels $Y_{label}$
   Output:
 2 Generated 3D CT $P_{ct}$, segmentation result $P_{label}$
   Hyperparameters:
 3 Total training epochs $N \in N^+$, weights $\alpha_1, \alpha_2 \in R$
   Training network:
 4 Train a network $f_\theta$ with 880 training samples and 100 validating
     samples.
 5 Initiallize network parameters $f_\theta \leftarrow f_{\theta,0}$
 6 for $i = 1, 2, \cdots N$ do
 7   |  $P_{ct}, P_{label} = f_\theta(X_{drr})$
 8   |  $L_{mse,\theta} = MSELoss(P_{ct}, Y_{ct})$
 9   |  $L_{bce,\theta} = BCELoss(P_{label}, Y_{label})$
10   |  $L_{loss,\theta} = \alpha_1 L_{mse,\theta} + \alpha_2 L_{bce,\theta}$ ($\alpha_1 = \alpha_2 = 1$)
11   |  updating parameters $\theta$ by minimization: $f_\theta \leftarrow f_{\theta,i}$
12 end
13 return $f_\theta$
```

## 3.2 Experimental setting

3.2.1 Training and testing data

Fifteen patient cases from the SPARE Challenge [46] were used to validate the proposed method. Each patient case contains a planning 4D CT with ten phases of 3D CT and segmentation labels of the tumors sited in the thoracic or upper-abdominal regions, with each 3D CT representing the anatomic structure at a specific respiratory phase.

Training CNN model requires X-ray projections, corresponding 3D CTs and tumor masks, which are not available in reality. As in previous studies, 3D CTs at different phases derived from 4D CT were utilized to generate digitally reconstructed radiographs (DRRs) as digital simulations of X-ray projections. First, all CT images and tumor labels were resampled to an isotropic resolution of 1×1×1 mm³ per voxel using cubic interpolation for CT and nearest neighbor interpolation for labels. Then, a PCA-MM is built with three principal components which are sufficient to describe most of the motion according to previous studies. Based on the PCA-MM, 1080 displacement vector fields (DVFs) were calculated by randomly sampling the PCA-MM coefficients. The corresponding CTs and tumor masks were obtained by deforming the reference CT and tumor masks with the sampled DVFs. A DRR was calculated for each CT at a random angle. A total of 1080 training samples were generated for each patient, of which 880 samples were used for model training and 100 samples for model validating. The remaining 100 samples were used for model testing.

Similar to previous studies[39], the input and output dimensions of the network were set to 128×128 and 128×128×128. CTs and corresponding DRRs were resized accordingly. Following the standard protocol of data pre-processing, scaling normalization was conducted for both CT and DRR, with their intensity normalized to the interval [0,1].

3.2.2 Compared methods and evaluation metrics

To benchmark the proposed method, a comparative study was conducted with three methods: PatRecon[20], X2CT[39] and TransNet [4]. While PatRecon and X2CT are two significant methods with available source code, TransNet is also employed as a comparative method as it innovatively utilizes segmentation as partial supervision. Though X2CT was initially proposed as population-based imaging model taking two orthogonal X-ray projection as input, it can be easily adapted to achieve patient-specific imaging based on one X-ray projection. The implementation from their publicly available code is adopted which does not change the network but uses the single-angle X ray twice.

As mentioned above, current methods are designed for fixed-angle CT reconstruction. Therefore, two types of experiments were conducted for evaluation. One was the angle-agnostic imaging experiments as discussed above and the other was the fixed-angle imaging experiments. For the angle-agnostic imaging experiments, the trained models are denoted as X2CT-R, TransNet-R and PatRecon-R. As for the fixed-angle imaging experiments, 1080 samples were generated at a fixed angle, of which 880 random samples were used for model training and 100 samples for model validating, with the remaining 100 samples for testing. The trained models were denoted as X2CT-F, TransNet-F and PatRecon-F. Theoretically, an imaging model trained with fixed-angle X-ray projections tends to outperform a model trained with angle-agnostic X-ray projections in fixed-angle imaging experiments. To illustrate the performance of the proposed method, we only trained the RT-SRTS on the angle-agnostic data set and directly tested the trained model on the fixed-angle data set.

The CT reconstruction results were evaluated in terms of the mean absolute error (MAE), mean square error (MSE), root mean square error (RMSE), structural similarity measure (SSIM) and peak signal-to-noise ratio (PSNR). In practice, MAE, MSE and RMSE are commonly used to evaluate the intensity accuracy between generated and ground truth CTs, whereas SSIM and PSNR mainly measure the visual quality of the generated CTs. As for the accuracy of tumor segmentation, the dice coefficient (DICE) and center of mass distance (COMD) are evaluated[4, 47]. The DICE represents the spatial overlap between the segmented and ground truth tumor volume and the COMD measures the distance between the center of masses. In general, lower MAE, MSE and RMSE values, and higher SSIM and PSNR values indicate a better reconstruction. Higher DICE value and lower COMD value indicate for a more precise tumor segmentation.

# 4 Results and discussion

## 4.1 Evaluation of reconstruction and segmentation performance

Approximately 15 hours were required to train the network for 100 epochs, and the best validation performance was achieved at approximately 90th epoch. The model with the best validation performance was subsequently used for testing. Notably, the proposed method is robust to the weights for the MSE and BCE losses and the fixed weights ($\alpha_1 = \alpha_2 = 1$) achieves consistently good performance across all test cases. In terms of inference speed, the trained network demonstrates efficient processing capabilities. It takes approximately 70 ms for the network to generate both the 3D CT and 3D tumor mask from a single X-ray projection. This processing time is significantly shorter than the maximum allowed delay time of 500 ms [3, 48]. Therefore, it is believed that the proposed method can be implemented in an online tracking workflow for VMAT.

The quantitative results of the CT reconstruction are presented in Figure 4. As demonstrated, RT-SRTS outperformed all other methods in terms of all metrics for the angle-agnostic experiments. For the fixed-angle imaging experiments, RT-SRTS also outperformed others in most metrics except for SSIM, albeit with a negligible difference in performance. The CT reconstruction results were qualitatively evaluated in Figure 5. All four methods could effectively capture tissue displacement caused by respiratory motion with similar tissue position and shape to the ground truth. However, X2CT fails to reconstruct acceptable images for angle-agnostic experiments, and generates very noisy volumes with many artefacts for fixed-angle experiments. In contrast, the results of TransNet are very blurry with missing anatomic details in both experimental settings. Although PatRecon achieves satisfactory reconstruction results in fixed-angle experiments, its performance deteriorates in angle-agnostic experiments. This is clearly demonstrated in the highlighted area in Figure 5(a), where PatRecon-R lacks discernible fine details and struggles to differentiate tissues at close distances. Furthermore, as is evident from Figure 5(b), the reconstruction results of PatRecon-R become blurred at some highlighted areas. In contrast, RT-SRTS and PatRecon-F could reconstruct the small anatomical structures better. In overall, for the compared methods, the results at fixed-angle setting are better than that at angle-agnostic setting. RT-SRTS obtain the clearest result with the finest details, and its results approach the ground truth more closely.

Furthermore, RT-SRTS can accurately segment the tumors with an average DICE of 0.96±0.03 and an averaged COMD of 0.40±0.18 mm. The 3D coordinates of the tumor are desired for tumor tracking[3] and the COMD is directly relevant to tumor localization accuracy. The achieved results in the COMD are relatively low compared with previously reported results [4] [24]. For example, a recent tumor tracking method reported a root mean square error of the tumor centroid less than 1.5 mm [24]. This further demonstrate the efficacy of the adopted MTL strategy and proposed modules. Figure 6 illustrates the accuracy of RT-SRTS in terms of tumor segmentation. For each slice, there is only a small deviation in the segmentation from the ground truth.

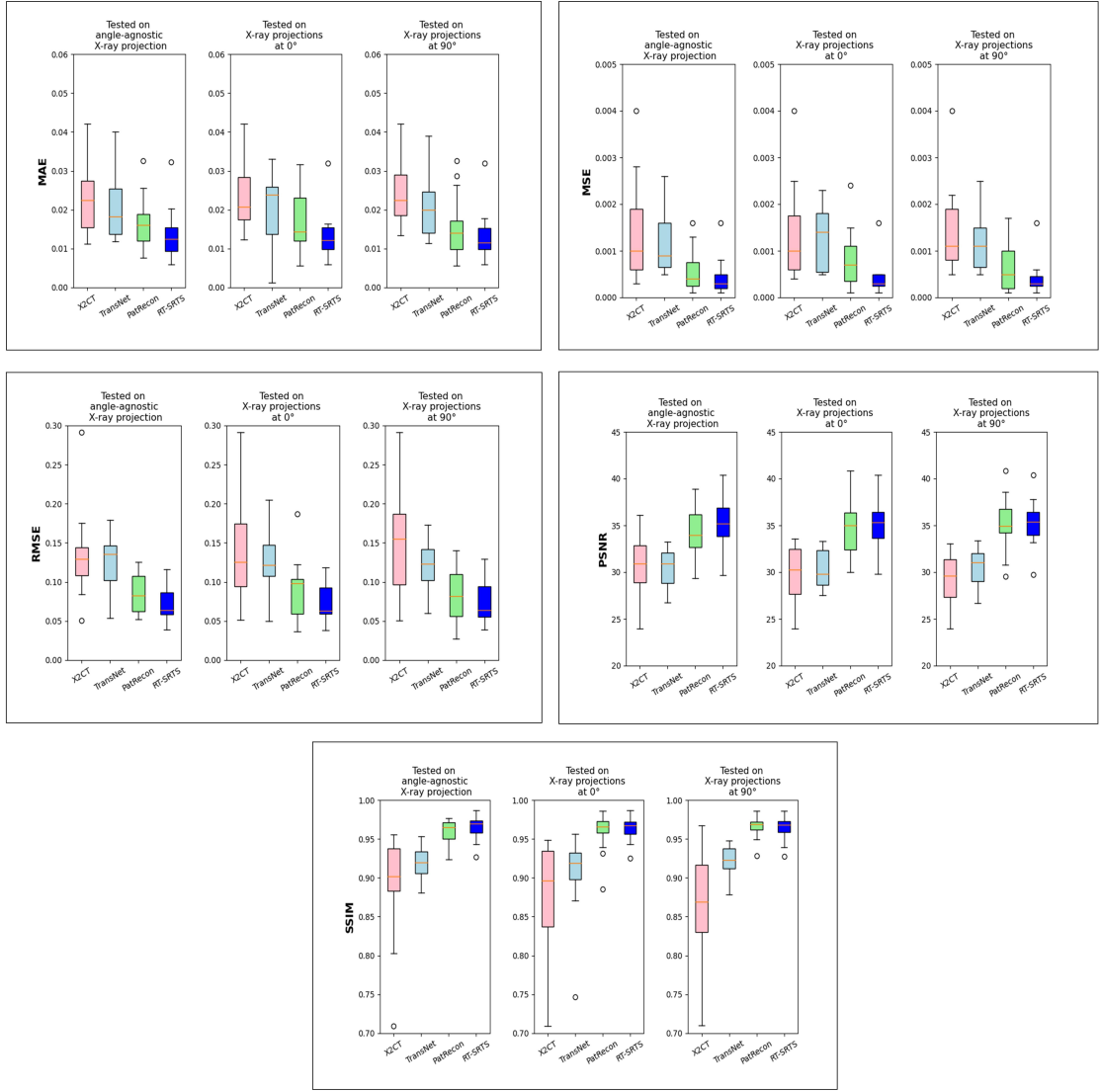

**Figure 4** Quantitative comparison of different methods. The five subplots depict the differences between RT-SRTS and the comparative methods in terms of MAE, MSE, RMSE, PSNR, and SSIM metrics. Each subplot presents the results for random angles and two fixed angles (0 degrees and 90 degrees) separately. The results of RT-SRTS for each case are shown in Appendix A.

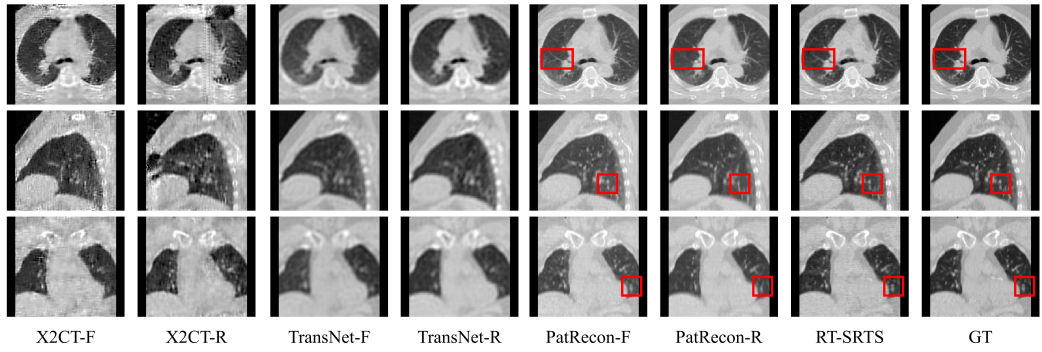

(a) Reconstruction from X-ray projection at the angle of 90°

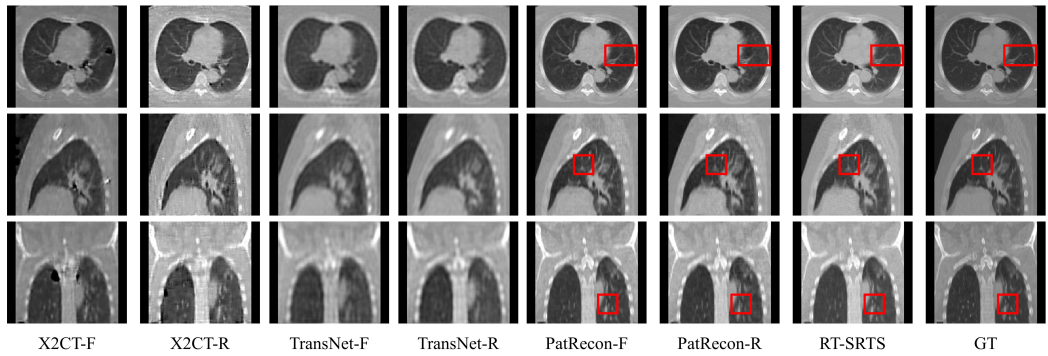

(b) Reconstruction from X-ray projection at the angle of 0°

**Figure 5** Visual comparison between the reconstruction results of different methods.

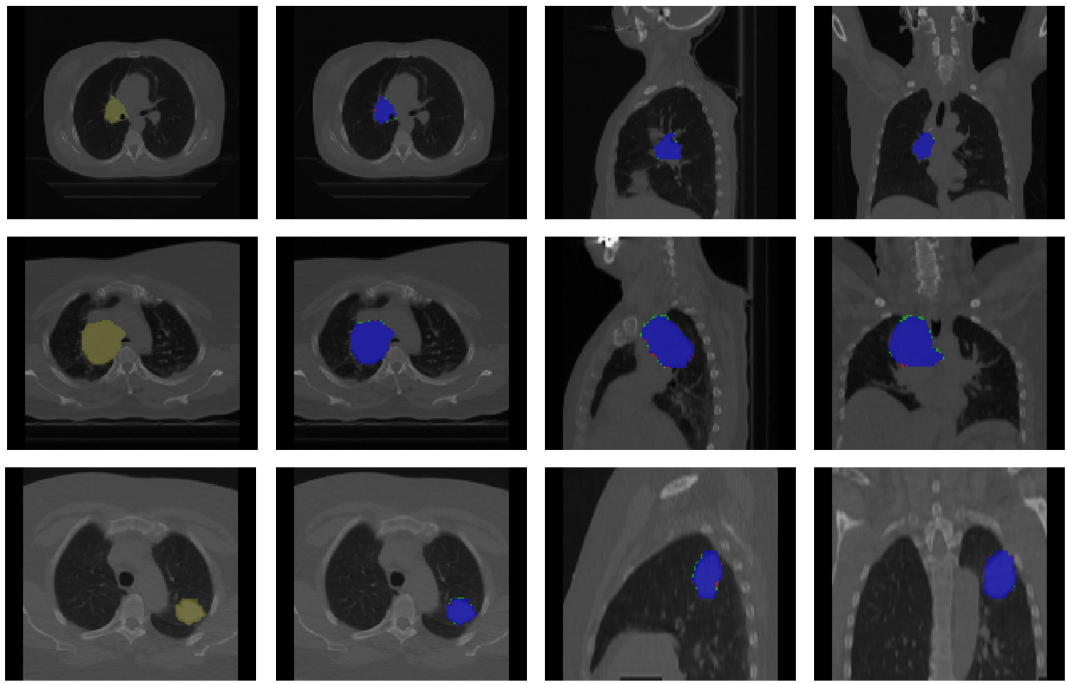

**Figure 6** Visualization of tumor segmentation results for three cases with one row for each case. While the first column shows the ground truth mask in the axial slices, the following three columns compare the segmentation results with the ground truth. Red, green and blue indicate false negative, false positive and true positive, respectively.

## 4.2 Evaluation of robustness to noise

The inevitable presence of artifacts (noise, scatter, etc.) in real X-ray projection causes appearance disparity between the simulated DRRs and real X-ray projections. To investigate the effect of noise on the performance of the proposed method, according to a previous study[4], the performance of RT-SRTS was also evaluated on noisy testing projection data by adding Gaussian noise with sigma ratio of 0.01, 0.02, and 0.05, whereas the model was trained on DRRs without adding noise. Table 1 presents the performance metrics of the model at various noise levels. Although there's a minor degradation in the reconstruction as the noise level increases, the results remain largely consistent. Similarly, the segmentation accuracy declines slightly for noisy projections, but the outcomes are comparable. As shown in Figure 7, there is only a minor difference between the reconstruction and segmentation results.

**Table 1.** Metric values of the 3D images generated by RT-SRTS from projection data added with Gaussian noise of varying sigma ratios.

| | 3D Reconstruction | | | | | Tumor Segmentation | |
|---|---|---|---|---|---|---|---|
| Ratio | MAE | MSE | RMSE | PSNR (dB) | SSIM | DICE | COMD (mm) |
| 0 | 0.013 | 0.0004 | 0.072 | 35.38 | 0.965 | 0.962 | 0.40 |
| 0.01 | 0.015 | 0.0005 | 0.083 | 34.48 | 0.952 | 0.942 | 0.46 |
| 0.02 | 0.016 | 0.0005 | 0.089 | 33.82 | 0.950 | 0.943 | 0.48 |
| 0.05 | 0.018 | 0.0007 | 0.099 | 32.74 | 0.938 | 0.941 | 0.56 |

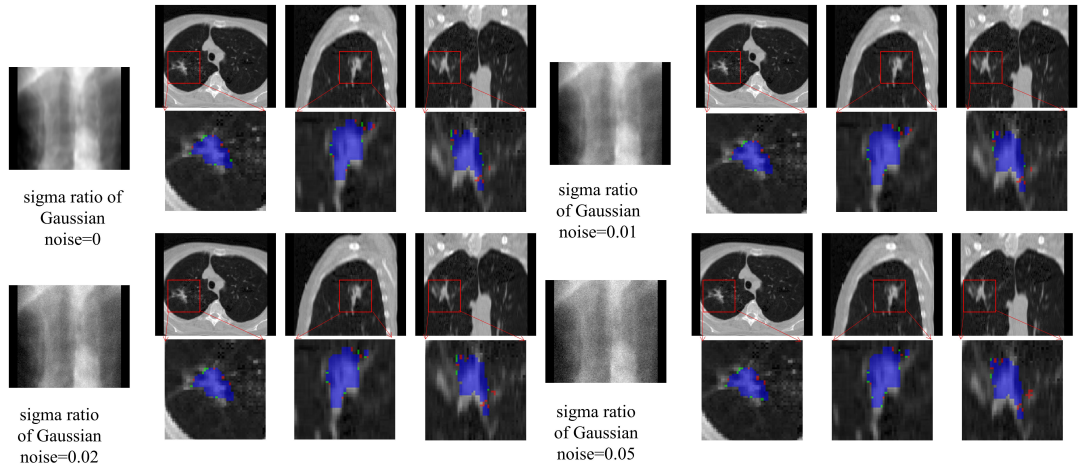

**Fig 7** Visualization of noisy projection data and their corresponding reconstruction and segmentation results.

### 4.3 Ablation Study

**Table 2. The results of the ablation study.**

| | | | 3D Reconstruction | | | | | Tumor Segmentation | |
|---|---|---|---|---|---|---|---|---|---|
| Segmentation branch | AEC | URE | MAE | MSE | RMSE | PSNR (dB) | SSIM | DICE | COMD (mm) |
| | | | 0.017 | 0.0011 | 0.099 | 33.57 | 0.908 | / | / |
| √ | | | 0.017 | 0.0007 | 0.090 | 34.50 | 0.957 | 0.943 | 0.41 |
| √ | √ | | 0.014 | 0.0005 | 0.075 | 35.18 | 0.964 | 0.948 | 0.46 |
| √ | √ | √ | 0.013 | 0.0004 | 0.072 | 35.38 | 0.965 | 0.962 | 0.40 |

The prototype of RT-SRTS was derived from the PatRecon method, with minor adjustments to the convolutional kernels. Building upon this baseline, we novelly incorporated the segmentation branch, as well as the AEC and URE modules. Ablation experiments were conducted to assess the effectiveness of these enhancements, and the results are shown in Table 2. As we can see, compared with the baseline (the first row), the integration of the segmentation branch not only achieves proficient tumor segmentation but also enhances the reconstruction performance, aligning with expectations from the multi-task learning theory. Furthermore, there is a significant improvement in all metrics by adding the AEC and URE modules. Specifically, the AEC module significantly improves reconstruction metrics such as PSNR, whereas the URE module contributes to a greater enhancement in segmentation DICE.

The AEC module not only leverages skip connections to transmit diverse scale features from the feature extractor to both the reconstruction and segmentation branches but also incorporates an attention mechanism. This mechanism facilitates the integration of feature information, allowing the reconstruction and segmentation branches to retain the detailed features. Consequently, this approach enhances the ability to reconstruct finer anatomy details. As shown in Figure 8, the network equipped with the AEC module demonstrates a significant improvement in the reconstruction details of skeletal structures. Quantitatively, the performance of reconstruction gains 0.68 dB in PSNR and 1% in SSIM. The URE module utilizes pixels with high confidence to rectify pixels with low confidence, thereby, improving the segmentation accuracy around the tumor boundary. This can be exemplified by comparing the segmentation results with/without the URE in Figure 9. Table 2 shows that the DICE increased from 0.948 to 0.962 through URE.

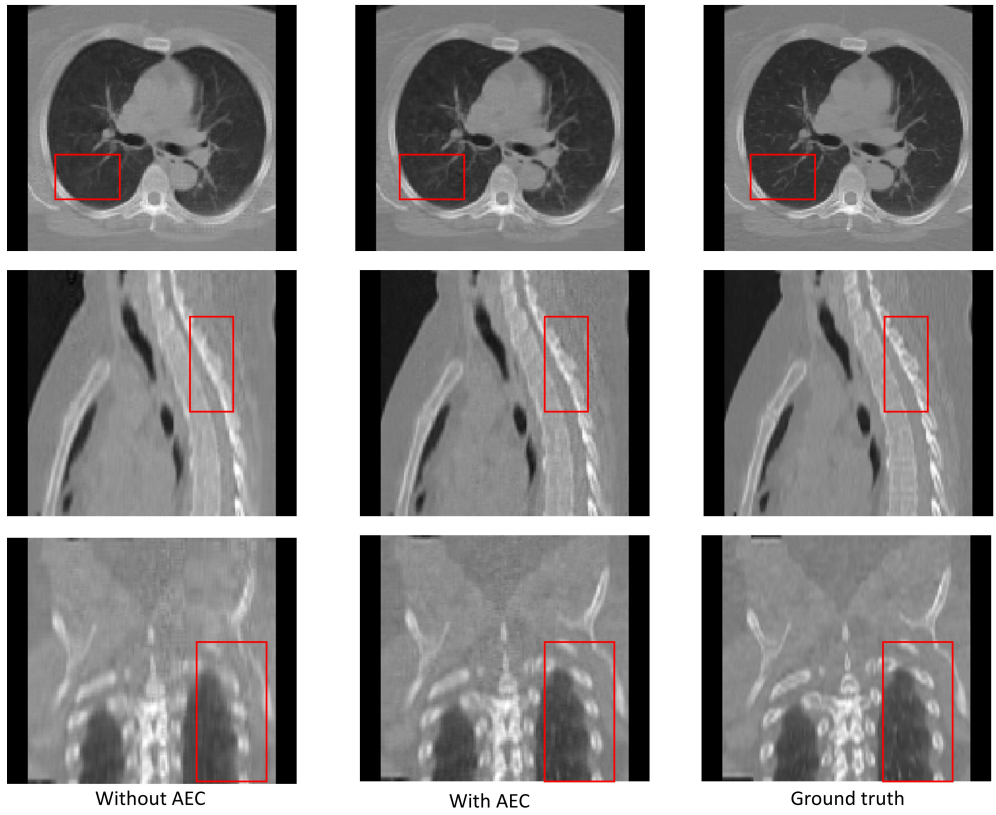

**Figure 8** Comparison of the reconstruction results w/wo AEC module.

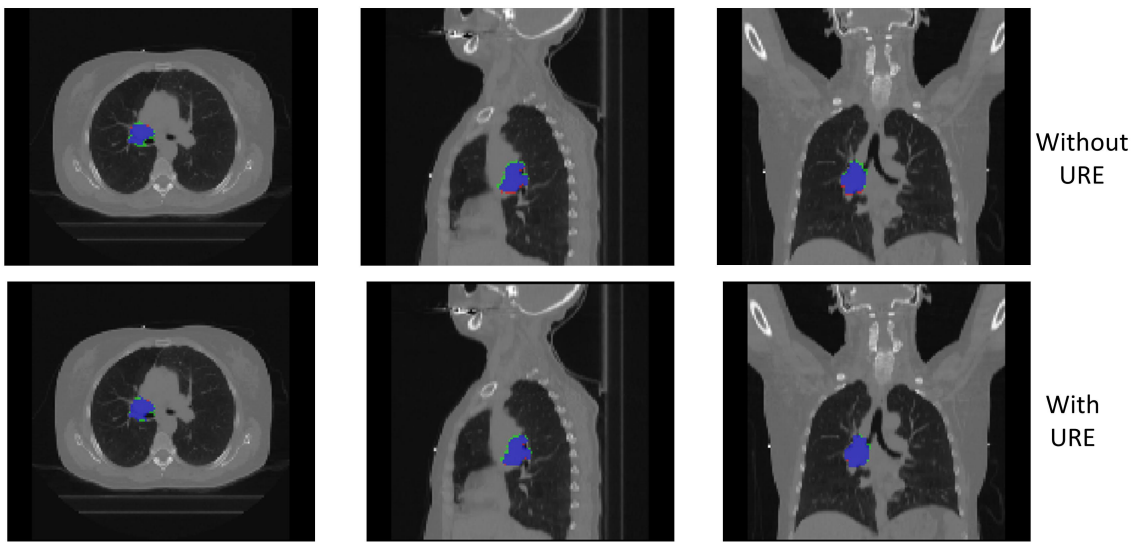

**Figure 9** Comparison of the segmentation results w/wo URE module. Red, green and blue indicate false negative, false positive and true positive, respectively.

## 5 Conclusion

In this study, a novel patient-specific CNN imaging method, RT-SRTS, is proposed which is able to achieve simultaneous real-time 3D reconstruction and tumor segmentation from one X-ray projection acquired at any angle. The RT-SRTS contains reconstruction and segmentation sub-networks with a shared representation network utilized to extract hierarchical semantic features from the input 2D X-ray projection. To further improve the performance, an attention enhanced calibrator (AEC) module was proposed to fuse hierarchical features with the imaging and segmentation branches more effectively, and an uncertain-region elaboration (URE) module was proposed to improve the segmentation accuracy at the tumor boundary. The effectiveness of RT-SRTS was evaluated in both fixed-angle and angle-agnostic imaging modes on 15 patient cases. Compared to X2CT, TransNet and PatRecon, RT-SRTS outperforms in terms of MAE, MSE, RMSE, PSNR, and SSIM metrics, demonstrating its capability to achieve superior 3D reconstruction. As for the tumor segmentation and localization, the obtained COMD of 0.40±0.18 mm compares favorably with recent tumor location methods. In addition, the inference time for RT-SRTS is 70±3 ms, which ensures the feasibility of real-time tracking. As RT-SRTS is the first CNN based CT imaging model that can fully meet the motion control requirement for VMAT radiotherapy, it serves as a crucial development for VMAT image guidance and has the potential to be implemented in an online tracking workflow to improve the treatment accuracy.

Due to the non-isotropic nature of the human anatomy, the quality of X-ray projection varies with the projection angle, which may directly affect the 3D reconstruction quality and tumor localization accuracy. Indeed, it has been noted that the performance of fixed-angle imaging or tumor localization varies with the projection angle for existing methods. For instance, it's commonly observed that performance tends to be superior at 90° (Anterior-posterior direction) compared to 0° (Left-right direction) [24, 49] because the X-ray projection at 0° exhibits finer contrast of soft tissue due to the thinner anatomy at that direction, whereas the opposite for X-ray projection at 90°. Regarding our angle-agnostic method, experimental findings indicate it is relatively insensitive to projection angles. In fixed-angle imaging experiments, the PSNR and SSIM achieved at 0° (35.2±5.6, 0.963±0.00023) are comparable to those obtained at 90° (35.3±5.4, 0.964±0.00022). This resilience to angle variations may be attributed to the fact that our RT-SRTS model was trained on X-ray projections with random angles, making it more robust to changes in projection angle. We intent to further investigate this issue in our future work.

Despite the excellent performance, this work still has some aspects need to be improved in next work. First, although RT-SRTS has been validated on 15 patient cases, testing with more data is warranted to validate its performance fully before potential clinical deployment. Second, the current emphasis on network design primarily revolves around optimizing the information flow between the reconstruction and segmentation branches. The performance could be further refined by enhancing other aspects of the network design. For instance, preliminary exploration indicates potential benefits in merging features with different receptive fields, a finding that merits more in-depth investigation. Finally, given the patient-specific nature of this imaging method, the requirement to train a unique imaging model for each patient may increase the workload of the clinic staff. Hence, it is imperative to explore automatic training methods and enhance training efficiency using techniques such as transfer learning or meta-learning.

**Appendix A The testing results on fifteen cases**

Table A1 The testing results of RT-SRTS on fifteen cases.

|  | | 3D Reconstruction | | | | | Tumor Segmentation | |
| --- | --- | --- | --- | --- | --- | --- | --- | --- |
|  | Tumor Size (cm$^3$) | MAE | MSE | RMSE | PSNR (dB) | SSIM | DICE | COMD (mm) |
| Case1 | 6.96 | 0.0153 | 0.0004 | 0.0734 | 33.8817 | 0.9512 | 0.97392 | 0.456 |
| Case2 | 0.119 | 0.0162 | 0.0006 | 0.0807 | 33.7997 | 0.9652 | 0.89075 | 0.155 |

| Case3 | 0.313 | 0.0087 | 0.0002 | 0.0701 | 37.7828 | 0.9762 | 0.91732 | 0.163 |
| Case4 | 2.470 | 0.0116 | 0.0003 | 0.0392 | 36.2180 | 0.9748 | 0.98867 | 0.155 |
| Case5 | 3.020 | 0.0145 | 0.0004 | 0.1089 | 35.1537 | 0.9267 | 0.96940 | 0.052 |
| Case6 | 5.252 | 0.0322 | 0.0016 | 0.1063 | 29.6960 | 0.9680 | 0.97569 | 0.122 |
| Case7 | 1.434 | 0.0086 | 0.0002 | 0.0923 | 37.4820 | 0.9737 | 0.92827 | 0.239 |
| Case8 | 12.70 | 0.0059 | 0.0001 | 0.0481 | 40.4013 | 0.9864 | 0.97337 | 0.220 |
| Case9 | 3.563 | 0.0125 | 0.0003 | 0.0630 | 35.1389 | 0.9663 | 0.98666 | 0.467 |
| Case10 | 26.90 | 0.0115 | 0.0003 | 0.0605 | 35.1817 | 0.9695 | 0.99229 | 0.422 |
| Case11 | 2.496 | 0.0142 | 0.0005 | 0.0562 | 33.5462 | 0.9497 | 0.96468 | 0.458 |
| Case12 | 1.838 | 0.0108 | 0.0002 | 0.0642 | 36.5285 | 0.9723 | 0.89475 | 1.546 |
| Case13 | 7.274 | 0.0097 | 0.0003 | 0.0389 | 35.3339 | 0.9714 | 0.98552 | 0.359 |
| Case14 | 6.459 | 0.0157 | 0.0005 | 0.1161 | 33.3711 | 0.9432 | 0.98191 | 0.443 |
| Case15 | 7.269 | 0.0090 | 0.0002 | 0.0627 | 37.2918 | 0.9727 | 0.99707 | 0.060 |

**Acknowledgments**

This work was supported by the Beijing Natural Science Foundation (7232340, L222034 and L222104), the National Natural Science Foundation of China (12375359), and the Fundamental Research Funds for the Central Universities.

**Declaration of Generative AI and AI-assisted technologies in the writing process**

Statement: During the preparation of this work the authors used ChatGPT3.5 in order to improve readability and language. After using this tool, the authors reviewed and edited the content as needed and takes full responsibility for the content of the publication.

**CRediT authorship contribution statement**

Miao Zhu: Methodology, Writing - Original draft, Data curation, Visualization, Validation. Qiming Fu: Investigation, Methodology, Writing - review & editing, Data curation, Visualization. Bo Liu: Conceptualization, Investigation, Writing – review &

editing, Funding acquisition, Project administration, Supervision. Mengxi Zhang: Conceptualization, Methodology. Bojian Li: Data curation, Investigation. Xiaoyan Luo: Writing – review & editing, Data curation. Fugen Zhou: Supervision.

**Declaration of competing interest**

Declaration of interest: The authors declares that there are no conflicts of interest.

**Data availability**

The primary patient data can be obtained at [SPARE Challenge – Image X Institute (sydney.edu.au)](). Our code will be available at [https://github.com/ZywooSimple/RT-SRTS]().